\documentclass{ieeeaccess}
\makeatletter
\renewcommand{\fnum@figure}{Fig. \thefigure}
\makeatother
\usepackage{cite}
\usepackage{balance}

\usepackage{amsfonts}
\usepackage[cmex10]{amsmath}
\usepackage{caption}
\usepackage{subcaption}
\usepackage{adjustbox}
\usepackage{amssymb}
\def\BibTeX{{\rm B\kern-.05em{\sc i\kern-.025em b}\kern-.08em
    T\kern-.1667em\lower.7ex\hbox{E}\kern-.125emX}}    
\usepackage[acronym]{glossaries}
\newacronym{OFDM}{OFDM}{orthogonal frequency division multiplexing}
\newacronym{CP}{CP}{cyclic prefix}
\newacronym{JCAS}{JCAS}{joint communication and  radar sensing}
\newacronym{RF}{RF}{radio frequency}
\newacronym{QAM}{QAM}{quadrature amplitude modulation}
\newacronym{MUSIC}{MUSIC}{multiple signal classification}
\newacronym{ESPRIT}{ESPRIT}{estimation of signal parameters via rotational invariance technique}
\newacronym{CS}{CS}{compressive sensing}
\newacronym{FPS}{FPS}{Fourier projection-slice}
\newacronym{LMMSE}{LMMSE}{linear minimum mean square error}
\newacronym{LS}{LS}{least square }
\newacronym{MMSE}{MMSE}{minimum mean square error}
\newacronym{CE}{CE}{channel estimation}
\newacronym{RMSE}{RMSE}{root mean square error}
\newacronym{TDD}{TDD}{Time Division Duplexing}
\newacronym{FDD}{FDD}{Frequency Division Duplexing}
\newacronym{UL}{UL}{Uplink}
\newacronym{DL}{DL}{Downlink}
\newacronym{BS}{BS}{Base Station}
\newacronym{MRC}{MRC}{Maximum Ratio Combining}
\newacronym{CSI}{CSI}{Channel State Information}
\newacronym{ACI}{ACI}{adjacent channel interference}
\newacronym[longplural={User Equipments}]{UE}{UE}{User Equipment}
\newacronym{3GPP}{3GPP}{3rd generation partnership project }
\newacronym{LTE}{LTE}{long term evolution}
\newacronym{AoA}{AoA}{Angle Of Arrival}
\newacronym{SFT}{SFT}{sparse Fourier transform}
\newacronym{ED}{ED}{Energy Detection}
\newacronym{ASK}{ASK}{Amplitude-shift keying}
\newacronym{PSK}{PSK}{Phase-shift keying}
\newacronym{ZCP}{ZCP}{Zadoff-Chu precoding}
\newacronym{ISM}{ISM}{industrial scientific and medical}
\newacronym{SNR}{SNR}{signal-to-noise ratio}
\newacronym{IDFT}{IDFT}{inverse discrete Fourier transform }
\newacronym{DFT}{DFT}{discrete Fourier transform }
\newacronym{IFFT}{IFFT}{inverse fast Fourier transform }
\newacronym{FFT}{FFT}{fast Fourier transform }
\newacronym{ISI}{ISI}{inter symbol interference}
\newacronym{ICI}{ICI}{inter carrier interference}
\newacronym{MIMO}{MIMO}{Multiple Input Multiple Output}
\newacronym{PAPR}{PAPR}{peak-to-average power ratio}
\newacronym{GLRT}{GLRT}{Generalized Likelihood Ratio Test}
\newacronym{i.i.d}{i.i.d}{independent and identically distributed}
\newacronym{SS}{SS}{Spectrum Sensing}
\newacronym{AWGN}{AWGN}{additive white Gaussian noise}
\newacronym{ZCT}{ZCT}{Zaddof-Chu transform}
\newacronym{HPA}{HPA}{high power amplifier}
\newacronym{RFI}{RFI}{Radio Frequency Interferences}
\newacronym{2D}{2D}{two-dimensional}
\newacronym{MSE}{MSE}{mean square error}
\newacronym{BER}{BER}{bit error rate}
\newacronym{RADAR}{RADAR}{RAdio Detection And Ranging}
\newacronym{RCS}{RCS}{radar cross section}
\newacronym{ITS}{ITS}{Intelligent Transportation Systems}   
\begin{document} 
\history{Date of publication xxxx 00, 0000, date of current version xxxx 00, 0000.}
\doi{10.1109/ACCESS.xxxx.DOI}

\title{An Efficient OFDM-Based Monostatic Radar Design for Multitarget Detection}
\author{
\uppercase{Mamady Delamou}\authorrefmark{1},
\uppercase{Guevara Noubir\authorrefmark{2}, 
\uppercase{Shuping Dang\authorrefmark{3} and
El Mehdi Amhoud}\authorrefmark{1}}
}
\address[1]{School of Computer Science, Mohammed VI Polytechnic University, Benguerir, Morocco}
\address[2]{Khoury College of Computer Sciences, Northeastern University,  Boston, USA}
\address[3]{Department of Electrical and Electronic Engineering, University of Bristol, Bristol BS8 1UB, UK}

 \tfootnote{This work was sponsored by the Junior Faculty Development program under the UM6P – EPFL Excellence in Africa
Initiative}
\markboth
{Author \headeretal: An Efficient OFDM-Based Monostatic
Radar Design for Multitarget Detection}
{Author \headeretal: An Efficient OFDM-Based Monostatic
Radar Design for Multitarget Detection}
\corresp{Corresponding author: Mamady Delamou (e-mail: mamady.delamou@um6p.ma).}
\begin{abstract}
In this paper, we propose a monostatic radar design for multitarget detection based on orthogonal-frequency division multiplexing (OFDM), where the monostatic radar is co-located with the transmit antenna. The monostatic antenna has the perfect knowledge of the transmitted signal and listens to echoes coming from the reflection of fixed or moving targets. We estimate the targets parameters, i.e., range and velocity, using a two-dimensional (2D) periodogram. By this setup we improve the estimation performance under the condition of low signal-to-noise ratio (SNR) using Zadoff-Chu precoding (ZCP) and the discrete Fourier transform channel estimation (DFT-CE). Furthermore, since the dimensions of the data matrix can be much higher than the number of targets to be detected, we investigate the sparse Fourier transform based Fourier projection-slice (FPS-SFT) algorithm and compare it to the 2D periodogram. An appropriate system parameterization in the industrial, scientific and medical (ISM) band of 77 GHz , allows to achieve a range resolution of 30.52~cm and a velocity resolution of 66.79~cm/s.
\end{abstract}
\titlepgskip=-15pt
\maketitle
\section{Introduction}
\label{sec:introduction}
\PARstart{W}{ireless} communication systems and radio sensing systems are two different engineering paradigms     that have evolved separately in the past. However, nowadays, with the ever-increasing need for radio resources~\cite{jouhari2022survey}, implementing these two technologies separately leads to an inefficient utilization of the available spectrum. Despite their differences, communication and radar detection systems share many common features, particularly in terms of signal processing and equipment \cite{Rahman2021EnablingJC}. This has led several researchers to investigate the implementation of a unified system merging both technologies~\cite{5776640,article12,2019arXiv190105558L,8828023,8828016}.
Several approaches have been proposed in previous years, among which one of the prominent ideas is the coexistence of communications and radar detection~\mbox{\cite{article123,6331681,8835700}}. By such a model, the radar and communication systems can be co-located and even physically integrated. However, they transmit two different signals that overlap in the time and/or frequency domains. In order to minimize the interference between them, both systems need to operate simultaneously by sharing the same resources in a cooperative way~\cite{Rahman2021EnablingJC}. Nevertheless, with this coexistence, managing interference becomes a challenging task \mbox{\cite{article123,7536903}}. In~\cite{article1234,article12345}, the authors demonstrate that WiFi or Zigbee signals could be used for object sensing as well, showing the possibility to exploit communication signals for sensing.\\
\indent The difficulty in mitigating congestion and interference has led researchers to think of another solution which merges the communication and radar subsystems in a single device, using exactly the unified spectral and hardware resources. Such a conception is called in the literature~\gls{JCAS} or radio frequency (RF) \mbox{convergence~\cite{8714410}}.~\gls{JCAS} is significantly different from the existing concepts and from the aforementioned coexisting communication-radar systems. Instead, by \gls{JCAS}, the same waveform is used for both communication and sensing \cite{huang2019dual,Bjorn9827892,mccormick2017simultaneous}. Given that~\gls{OFDM} has been widely adopted in contemporary mobile communication standards~\cite{amhoud2021ofdm,zerhouni2021filtered}, employing~\gls{OFDM} waveform for detection/radar purposes has attracted increasing interest in recent years\cite{6875584,5073387,Rahman2021EnablingJC,9296833,amhoud2017concatenation}. While the communication receiver needs to perform the~\gls{CE} and then decodes the transmitted data, the radar only needs to apply a detection algorithm on the reflected signal in order to estimate the range and velocity of targets.\\
\indent Nowadays, several detection algorithms exist, such \mbox{as~\gls{MUSIC}},~\gls{ESPRIT} \cite{7935499,9166743}, \mbox{~\gls{CS} \cite{davenport_duarte_eldar_kutyniok_2012,4524050}}, as well as \mbox{matrix pencil \cite{157226}.} In this work we adopt the periodogram method for a \gls{2D} signal. The periodogram algorithm is mainly computed using \gls{DFT} and \gls{IDFT}. Ranges and velocities are contained in channel characteristics, which means that a precise estimation of these parameters by the radar system requires accurate channel state information. However, in the low~\gls{SNR} region, achieving an error-free channel estimate is a daunting task.\\
\indent To overcome this challenge, we introduce the \mbox{\gls{DFT}-\gls{CE}} approach to reduce the noise level in the estimation. Additionally, we use \gls{ZCP} and show that it can improve the estimation performance. Furthermore, even though using \gls{FFT} and \gls{IFFT} to compute the periodogram is generally efficient, \gls{FFT}/\gls{IFFT} do not take into account the signal structure whereas a plethora of algorithms can be even faster by considering the signal sparsity \cite{4472244,8519339,inbooknearly,6736670}. Consequently, we investigate the \gls{FPS}-\gls{SFT} \mbox{\cite{8519339}} and discuss the trade-off to be taken between the estimation time and the estimation accuracy.\\
In sum, our main contributions in this paper are summarized as follows:
\begin{itemize}
    \item By precoding the \gls{OFDM} symbols with a Zadoff Chu code matrix, we first reduce the \gls{PAPR} of the \gls{OFDM} signal and then, decrease the estimation error of ranges and velocities in the low \gls{SNR} region.
    \item Furthermore, \gls{DFT}-\gls{CE} is a channel estimation algorithm used to reduce the amount of noise in the frequency bin single-tap channel estimate. In this work, we adapt it to filter false positive targets. By combining it with \gls{ZCP}, we come up with better targets' range and velocity estimates.
    \item Finally, knowing that in practice the number of effective targets is smaller compared to the whole frame dimension, the signal is sparse in frequency domain. We take advantage of this characteristic and apply the~\gls{FPS}-\gls{SFT} to reduce the computational complexity along with the number of signal samples needed for estimation. By comparing it with the \gls{2D} periodogram, we observe a compromise between the execution time and the accuracy in the low SNR region.
\end{itemize}

The remainder of this paper is organized as follows. In Section II, we introduce the problem statement, the system model, and the periodogram based radar processing. Afterwards, we discuss the estimation improvements, using the \gls{DFT}-\gls{CE} combined to~\gls{ZCP} and applying the~\gls{FPS}-\gls{SFT} approach for reducing the computational complexity associated with the \gls{2D} periodogram processing in Section III. Simulation results and discussions are presented in Section IV. Finally, Section V concludes the work and sets forth our perspectives.
\section{Problem Statement and system model}
\subsection{Problem Statement}
\begin{figure}[t] 
\centering
\includegraphics[width=3.5in]{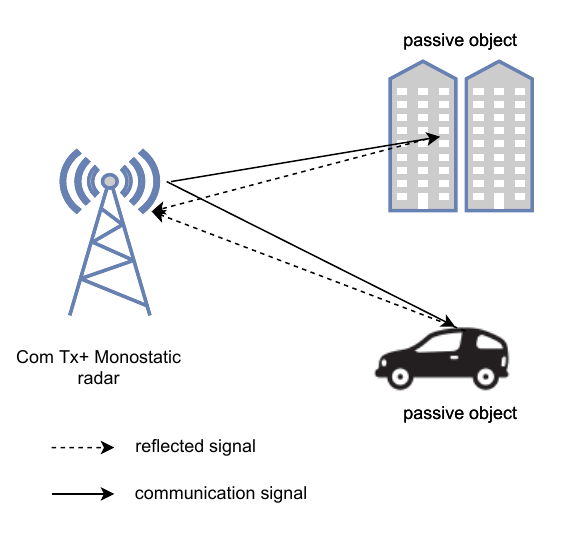}
\caption{System model of the monostatic radar detection.}
\label{system_model}
\end{figure}

We consider a wireless communication system consisting of a communication antenna Tx co-located with a monostatic radar as depicted in Fig. \ref{system_model}. In the downlink, the signal emitted from the communication subsystem, is known to the radar, and is reflected by a certain number of targets characterized by their ranges and velocities.~\gls{OFDM} is one of the leading technologies used in contemporary wireless communication systems. Considerable attention has been given to~\gls{OFDM} for its performance advantages, such as its ability to mitigate \gls{ICI} and \gls{ISI} by making a suitable use of a~\gls{CP}, as well as its robustness against frequency selective fading in addition to its efficient spectral utilization. In this regard, we adopt \gls{OFDM} as the multiplexing scheme since using a single signal for communication and sensing is strongly dependent to the data structure \cite{6875584}. The transmitted signal consists of pilots used for channel estimation and net information. \\
\indent It is important to mention that channel estimation needs to be performed for both the communication and the radar subsystems, but in different ways and for different purposes. At the communication receiver, as usual, the received frequency-domain pilots are used to perform channel estimation using either \gls{LS}-\gls{CE},  \gls{MMSE}-\gls{CE}, or any other channel estimation algorithms. Depending on the type of pilots arrangements, either frequency or time domain interpolation (or both) can be performed to infer the channel on the non-pilot subcarriers of each symbol \cite{Braun2014OFDMRA,khelouani2021ufmc}. At the radar, which we treat in this work, the entire grid of the transmitted signal is used to perform the channel estimation~\cite{Braun2014OFDMRA,khelouani2021ufmc} since it knows the transmitted frames. The problem consists of estimating the channel in order to efficiently approximate the ranges and velocities of the targets using the reflected signals as shown in \mbox{Fig. \ref{system_model}}, and further, improve channel estimation. In this work, we focus on the downlink for simplicity, whereas most results and insights can be easily extended to the uplink. We also hypothesize that the interference between the radar and the communication antenna is negligible.

\subsection{System Model}
\begin{figure*}[t] 
\centering
\includegraphics[width=6in]{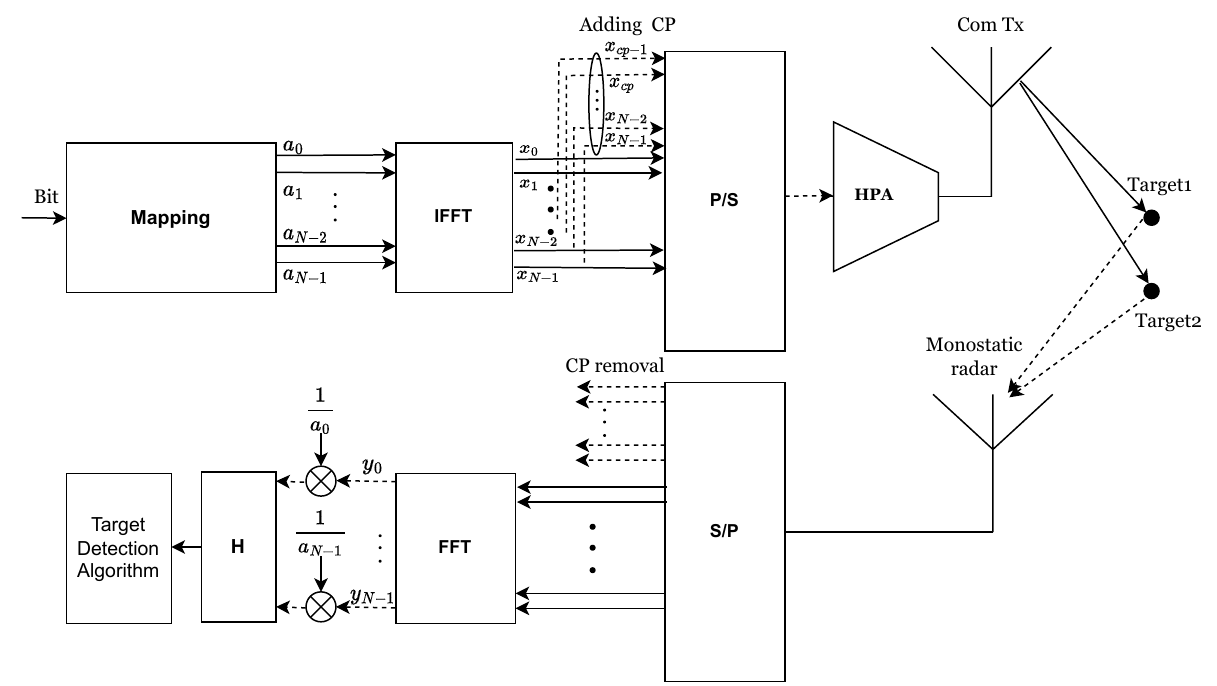}
\caption{OFDM-based monostatic radar transmission and detection process.}
\label{OFDM_JCAS_Process}
\end{figure*}

In this section, we describe the communication and radar channel model and explain how the radar performs the target  parameter estimation. We consider a total bandwidth $B$ which can be divided into $N$ small bands with central frequencies $f_0$,$f_1$\dots$f_{N-1}$ such that $\Delta f=\frac{B}{N}$. An \gls{OFDM} symbol is a packet of $N$ modulated data transmitted at the same time on $f_0$,$f_1$\dots$f_{N-1}$. The \gls{OFDM} symbol duration $T$ is thus given by $T=\frac{1}{\Delta f}$. After modulating bits by \gls{QAM}, \gls{IFFT} is applied to the \gls{QAM} data symbols to generate \gls{OFDM} waveforms in the time domain. Subsequently, a~\gls{CP} is added between consecutive symbols to mitigate~\gls{ISI}. Several~\gls{OFDM} waveforms are summed up to obtain an~\gls{OFDM} frame. The signal goes through a \gls{HPA} and the communication antenna Tx transmits it. The channel is a multi-path channel, to which an~\gls{AWGN} is added. At the radar, inverse operations are executed. First, the~\gls{CP} is removed, and then \gls{FFT} is performed on the~\gls{OFDM} bandpass signals. Finally, after the spectral division, targets detection algorithm is applied. The holistic process of transmission and detection is depicted in \mbox{Fig. \ref{OFDM_JCAS_Process}}.\\
\indent We assume that the complex symbols $\{a_{k,l}\}$ are generated after~\gls{QAM} modulation. Taking \gls{IFFT} on the zeroth \gls{OFDM} symbol $a_{k,0}$, the~\gls{OFDM} symbol sampled at sampling time $T_0=\frac{T}{N}$ can be represented as \cite{Braun2014OFDMRA}\mbox{:}
\begin{equation}\label{OFDM_sym}
x\left[n\right]=x(n T_{0})=\sum^{N-1}_{k=0} a_{k, 0} e^{j 2 \pi \frac{n k}{N}}, \quad\quad
0\le n \le N-1,
\end{equation}
which is the \gls{IFFT} of the \gls{QAM} symbols.\\
\indent Taking into account the \gls{CP} transmission time $T_{cp}$, an \gls{OFDM} symbol transmission time $T_s$ becomes $T_s = T+T_{cp}$. In terms of the number of symbols, we have $N_s=N+N_{cp}$, where $N_{cp}$ is the number of complex symbols transmitted during $T_{cp}$. Assuming that an \gls{OFDM} frame is composed of $M$ \gls{OFDM} symbols, the transmitted signal can be represented as:\\
  \begin{equation}\label{frame_resp}
      x[n]=\sum^{M-1}_{l=0} \sum^{N-1}_{k=0} a_{k,l} 
      e^\frac{{j2\pi k(n-lN_{s})}}{N}.
 \end{equation}
 
 \subsubsection{Radar channel model}
 We consider a baseband signal $x(t)$ with carrier frequency $f_c$. The transmitted passband  signal is $x_{pb}(t)=x(t)e^{j2\pi f_ct}$. For a given reflecting target at distance $d$ from the transmitter and moving at velocity $v$, the received passband signal is impacted by \mbox{\cite{Braun2014OFDMRA}:}
 \begin{itemize}
     \item Attenuation factor $g$ which depends on distance $d$, \gls{RCS} $\sigma_{RCS}$, carrier frequency $f_c$ and speed of light $c$, and by using the Friis equation of transmission, we obtain:
     \begin{equation}\label{attenuation}
         g=\sqrt{\frac{c\sigma_{RCS}}{(4\pi)^3d^4f_c^2}}.
     \end{equation}
     
     \item Signal delay $\tau$ caused by the round-trip, and $\tau=\frac{2d}{c}$.
     \item Doppler-Shift $f_D$ caused by the velocity of the target, and $f_D=\frac{2v}{c}f_c$.
     \item Random rotation phase $\varphi$ introduced when hitting the target.
     \item \gls{AWGN} $z(t)$, and \( z(t) \sim \mathcal{N}(\mu,\,\sigma^{2}) \).
 \end{itemize}
  
 Assuming a total number of $N_t$ reflecting moving targets, and taking into account all the previous constraints, received passband signal $y_{pb}(t)$ is written as follows \cite{braun2010maximum,khelouani2021ufmc}\mbox{:}
 
\begin{equation}\label{equa1}
    y_{pb}(t)=\sum^{N_t-1}_{p=0} g_p x(t-\alpha_p (t) )
    e^{j2\pi f_c(t-\alpha_p (t) )}e^{j\varphi_p}
    + z_{pb}(t),
\end{equation}
 where $\alpha_p(t)=2(\frac{ d_p}{c}+\frac{v_p}{c}t)=\tau_p+B_pt$ and $B_p=2\frac{v_p}{c}$.\\
 \indent At the radar, the received baseband signal $y(t)$ is obtained by demodulating $y_{pb}$(t) as $y(t)=y_{pb}(t)e^{-j2\pi f_ct}$. Thereafter, (\ref{equa1}) can be rewritten as
 \begin{equation}\label{chan_resp}
      y(t)=\sum^{N_t-1}_{p=0} g_px(t-\alpha_p(t)) 
      e^{-j2\pi f_c \tau_p t}      
      e^{j2\pi f_{D_p} t} 
      e^{j\varphi_p}
      + z(t).
 \end{equation}
 Eq. (\ref{chan_resp}) is the received signal containing attenuation $g_p$, channel delays $\tau_p$, Doppler effects $f_{D_p}$ and time-scale factor $B_p$. In essence, $y(t)$ is the filtered signal of $x(t)$ with the channel \mbox{impulse:}
 \begin{equation}
h(t)=\sum_{p=0}^{N_{t}-1} g_p e^{-j 2 \pi f_{c} \tau_p} e^{j 2 \pi f_{D_p}t} e^{j\varphi_p} \delta\left((1-B_p)t-\tau_p\right),
\end{equation}
where $\delta(t)$ is the Dirac function. A discrete-time counterpart considering a sample time $T_0$ is given \mbox{by}
\begin{equation} \label{discret_ch}
\begin{aligned}
h[n]=h(n T_0)=\sum_{p=0}^{N_{t}-1} g_p e^{-j 2 \pi f_{c} \tau_p} e^{j 2 \pi f_{D_p}n T_{0}}  e^{j\varphi_p} \\ \times\delta\left[(1-B_p) n-\tau_p / T_{0}\right].
\end{aligned}
\end{equation}
The discrete form of (\ref{chan_resp}) can thus be written \mbox{as}
\begin{equation}\label{conv}
y[n]=x[n] \otimes h[n]+z[n],
\end{equation}
 where $\otimes$ is the convolution operation.
 
Knowing that $T_0=\frac{1}{\Delta f N}$, the discrete-time counterpart of (\ref{chan_resp}) can be obtained by introducing (\ref{discret_ch}) and (\ref{frame_resp}) in (\ref{conv}) \cite{khelouani2021ufmc}: 
 
\begin{equation}\label{disc_OFDM_frame}
  \begin{aligned}
y[n]=\sum^{N_t-1}_{p=0} \sum^{M-1}_{l=0} \sum^{N-1}_{k=0}  g_p a_{k,l}
      e^{j2\pi \frac{kn}{N}}
      e^{-j2\pi k \Delta f \tau_p}
      e^{\frac{-j 2 \pi k B_p n}{N}} \\ 
      \times e^{\frac{j 2 \pi k l N_{s}}{N}} 
      e^{\frac{-j 2 \pi f_{D_{p}} n}{N \Delta f}}
      e^{j\varphi_p}
      + z[n].
    \end{aligned}
 \end{equation}
 
The signal for the $l$th \gls{OFDM} symbol is expressed as
\begin{equation}\label{discr_frame}
\begin{gathered}
y[n,l]=\sum^{N_t-1}_{p=0} \sum^{N-1}_{k=0}  g_p a_{k,l}
      e^{j2\pi \frac{k(n-lN_s)}{N}}
      e^{-j2\pi k \Delta f \tau_p} \\
       \times e^{\frac{-j 2 \pi k B_p (n-lN_s)}{N}} 
      e^{\frac{j 2 \pi k l N_{s}}{N}} 
      e^{\frac{-j 2 \pi f_{D_{p}} (n-lN_s)}{N \Delta f}}
      e^{j\varphi_p}
      + z[n].\\
      =\sum^{N_t-1}_{p=0} \sum^{N-1}_{k=0}  g_p a_{k,l}
      e^{j2\pi \frac{kn}{N}}
      e^{-j2\pi k \Delta f \tau_p}\\ 
      \times e^{\frac{-j 2 \pi k B_p n}{N}} 
      e^{\frac{j 2 \pi k B_p lN_s}{N}} 
      e^{\frac{-j 2 \pi f_{D_{p}} n}{N \Delta f}}
      e^{\frac{j 2 \pi f_{D_{p}} lN_s}{N \Delta f}}
      e^{j\varphi_p}
      + z[n].
\end{gathered}
\end{equation}
According to \cite{Braun2014OFDMRA}, a large sub-carrier distance heavily alleviates the de-orthogonalizing effect of a frequency offset. Therefore, it must be ensured that $\Delta f$ is larger than the Doppler shift caused by the object with the maximum relative velocity $v_{max}$, i.e., 
\begin{equation}\label{vmax}
    v_{max} \ll \frac{c\Delta f }{2f_c},
\end{equation}
which depends on the wave parameterization. \\
\indent Our system parameterization satisfies (\ref{vmax}), and, consequently, ($e^{\frac{-j 2 \pi k B_p n}{N}},e^{\frac{j 2 \pi k B_p lN_s}{N}}, e^{\frac{-j 2 \pi f_{D_{p}} n}{N \Delta f}}) \longrightarrow (1,1,1)$. Therefore, (\ref{discr_frame}) can be reduced to a simpler form infra: 
\begin{equation}
\begin{gathered}
y[n,l]=\sum^{N_t-1}_{p=0} \sum^{N-1}_{k=0}  g_p a_{k,l}
 e^{j2\pi \frac{kn}{N}}
 e^{-j2\pi k \Delta f \tau_p}
 e^{\frac{j 2 \pi f_{D_{p}} lN_s}{N \Delta f}}
 e^{j\varphi_p}\\
       + z[n].
\end{gathered}
\end{equation}

\subsubsection{Radar processing}

\begin{figure}[t!]
\centering
\includegraphics[width=3.2in]{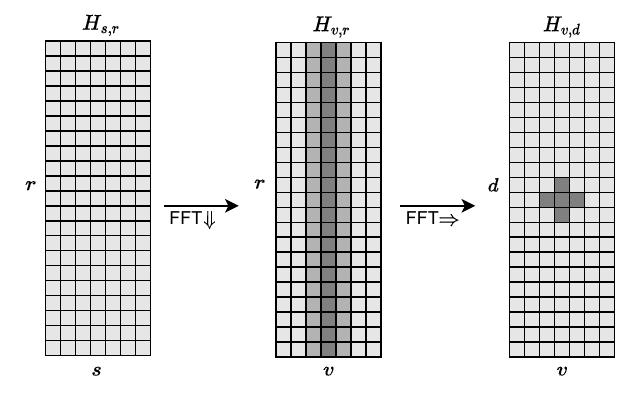}
     \caption{2D periodogram computation process: The vertical $\Downarrow$ and horizontal $\Rightarrow$ arrows indicate FFT over columns and rows, respectively.}
     \label{periodogram_process}
 \end{figure}

 The main purpose of using the \gls{CP} is to avoid \gls{ISI}. For that, the maximum channel delay $\tau_{max}$ should be less than the cyclic prefix transmission time, i.e.,    
 \begin{equation}\label{cylic_prefix_condition}
    \max_{0\le p \le N_t-1}\{\tau_p\}\le T_{cp}.
 \end{equation}
 We regulate that range $d_{un}$ (respectively a velocity $v_{un}$) is unambiguous if two targets positioned at $d$ and $d+d_{un}$ (respectively at moving velocities $v$ and $v+v_{un}$) cannot be distinguished~\cite{Braun2014OFDMRA}, given
 \begin{equation}\label{unambiguity}
  d_{un}=\frac{c}{2\Delta f},~\mathrm{and}~v_{un}=\frac{c}{2 f_c T_s}.
 \end{equation}
A distance $\Delta d$ (respectively velocity $\Delta v$) is called the radar resolution if it is the lowest distance (respectively velocity) such that two targets positioned at $d$ and $d+\Delta d$ (respectively moving at velocities $v$ and $v+\Delta v$) can still be distinguished~\cite{Braun2014OFDMRA}, i.e., 
\begin{equation}\label{radar_resolution}
      \Delta d=\frac{c}{2 N \Delta f}, ~\mathrm{and}~\Delta v=\frac{c}{2 M f_c T_s}.
 \end{equation}
 
\begin{figure}[t!] 
\centering
\includegraphics[width=3.5in]{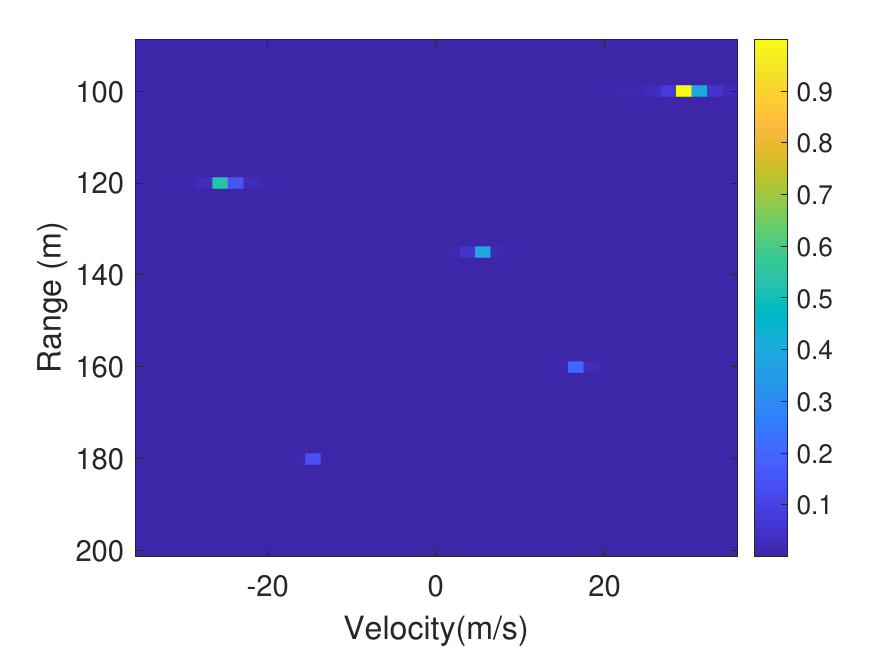}
\caption{Radar image map for 5 targets within the ranges of \mbox{100 m}, \mbox{120 m}, \mbox{135 m}, \mbox{160m}, \mbox{180 m} and moving at the velocities of \mbox{30 m/s}, \mbox{-25 m/s}, \mbox{5 m/s}, \mbox{17 m/s}, \mbox{-15 m/s}, given \mbox{\gls{SNR} = 5 dB}.}
\label{range_Doppler_map}
\end{figure}

The first step of the estimation process is the spectral division. Since the radar knows the transmitted frames, the channel information can be retrieved by calculating the ratio of the whole received signal over the whole transmitted one, which results in the \gls{LS}-\gls{CE} estimate of the channel. Then, estimation matrix $\boldsymbol{H}$ has entries as follows: \\
\begin{equation}\label{est_matrix}
\begin{aligned}
h_{k,l} &= \frac{y_{k,l}}{a_{k,l}} 
 = \sum^{N_t-1}_{p=0} b_{p}
  e^{j2\pi \frac{lN_sf_{D_p}}{N\Delta f}}
  e^{-j2\pi k\Delta f \tau_p}
  e^{j\Phi} 
 + \tilde{z}_{k,l},
\end{aligned}
\end{equation}
where $h_{k,l}$, $y_{k,l}$, $a_{k,l}$, and $z_{k,l}$ are the entry   at the $k$th row and the $l$th column in matrix $\boldsymbol{H}$, received frame matrix $\boldsymbol{Y}$, transmitted frame matrix $\boldsymbol{A}$ and noise matrix $\boldsymbol{\tilde{Z}}$, respectively; $\Phi$ is the phase obtained after the element-wise division. A periodogram is an estimate of the spectral density of a signal. Since in our case $\boldsymbol{H}$ is a two-dimensional signal, the corresponding periodogram can also be written as \cite{8714410}
\begin{equation}\label{periodogram}
P(s, r)=\frac{1}{NM}\left|\sum_{k=0}^{N^{\prime}-1}\left(\sum_{l=0}^{M^{\prime}-1}h_{k, l}w_{k, l} e^{-j 2 \pi \frac{l s}{R^{\prime}}}\right) e^{j 2 \pi \frac{k r}{S^{\prime}}}\right|^2,
\end{equation}
 where  $r=0, \ldots, N'-1 \quad$ and $\quad s=\left\lfloor\frac{-M'}{2}\right\rfloor, \ldots,\left\lfloor\frac{M'}{2}\right\rfloor\mbox{-1}$.
 Here, $\left\lfloor \frac{M'}{2} \right\rfloor$ indicates the floor of $\frac{M'}{2}$,  and the negative values of $r$ allow estimating negative velocities.
\begin{figure}[t!] 
\centering
\includegraphics[width=3.5in]{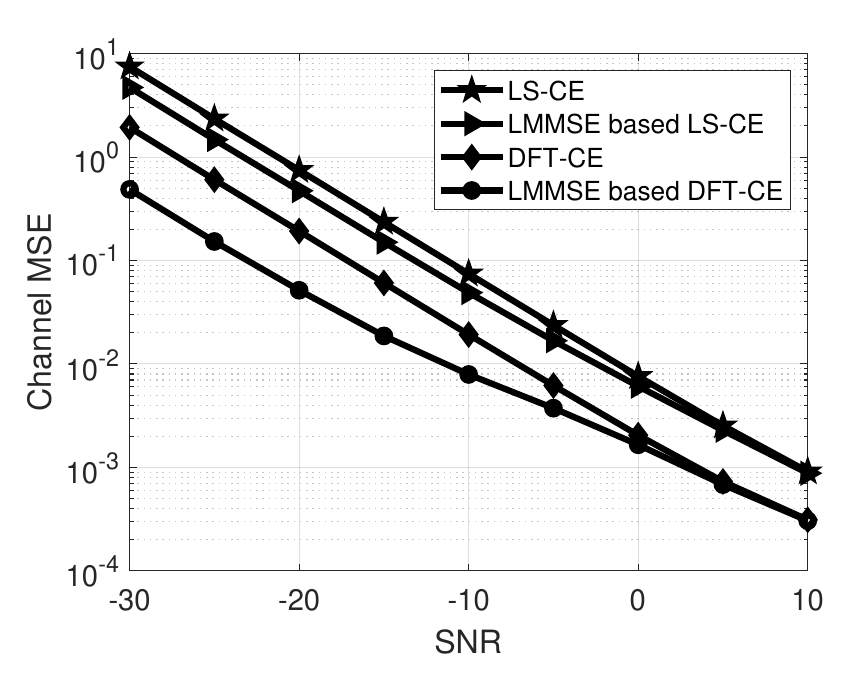}
\caption{Channel MSE for LS-CE, LMMSE based LS-CE, DFT-CE, and LMMSE based DFT-CE.}
\label{MSE_comparison}
\end{figure}
The problem here can be formulated as finding the optimal $s$ and $r$ in $\boldsymbol{P}$, by which the dominant frequencies are present. Those frequencies will represent the reflection points which are effective targets. In (\ref{periodogram}), $w_{k,l}$  is the value at the $k$th row and $l$th column in matrix $\boldsymbol{W}$. It is a window function that reduces the side-lobe levels of each dominant frequency. $N'$ and $M'$ are the extended values of $N$ and $M$ such that $N'\ge N$ and $M'\ge M$.  $N'$ and $M'$ can improve the precision of the estimation, but do not have any effect on the radar resolution. As such, (\ref{periodogram}) is equivalent to taking an $M'$-\gls{FFT} of each column of $\boldsymbol{H}$ then an $N'$-\gls{IFFT} of each row of the previously resulting matrix as depicted in Fig.~\ref{periodogram_process}. Consequently, (\ref{periodogram}) outputs dominant peaks, where targets are supposed to be located. Due to the whiteness of the noise, the detection threshold is equal to $\sigma^{2}\ln(P_{fa})$ \cite{Braun2014OFDMRA}. More explicitly, as established in (\ref{hypo}), any point \mbox{($s$, $r$)} such that $P(s,r) \ge \sigma^{2}\ln(P_{fa})$ is considered as a target, otherwise it is regarded as a noise, i.e.,  a false target, where $P_{fa}$ is the desired probability of false alarm. Mathematically, we have
\begin{equation}
P(s,r) \begin{cases} 
\ge \sigma^{2}\ln(P_{fa}) , & \text{target}\\ 
\le \sigma^{2}\ln(P_{fa}) , & \text{noise~~only}
\end{cases}
\label{hypo}.
\end{equation}

Once we obtain the list of estimates $\hat{s}$ and $\hat{r}$, the corresponding targets range and velocity values are deduced as follows: 
\begin{equation}\label{estimates}
    \hat{d}_p=\frac{c\hat{s}_p}{2N'\Delta f}, ~\mathrm{and}~\hat{v}_p=\frac{c\hat{r}_p}{2f_c M'T_0}. 
\end{equation}
Fig.~\ref{range_Doppler_map} shows a radar range-Doppler map, also called a radar image, of 5 targets at the ranges of \mbox{100 m}, \mbox{120 m}, \mbox{135 m}, \mbox{160 m}, \mbox{180 m}, moving at the velocities of \mbox{30 m/s}, \mbox{-25m/s}, \mbox{5 m/s}, \mbox{17 m/s}, \mbox{-15 m/s} when \mbox{\gls{SNR} = 5 dB}. As shown in this figure, the first target at the range of \mbox{100 m} and the velocity of \mbox{30 m/s} moving away from the transmitter, which is the closest to the transmitter and is well detectable (the top right target). The last target, the farthest at the range of \mbox{180 m} and the velocity of \mbox{-15 m/s} moving toward the transmitter is slightly detectable (the bottom left target).
\begin{figure}[t!] 
\centering
\includegraphics[width=3.5in]{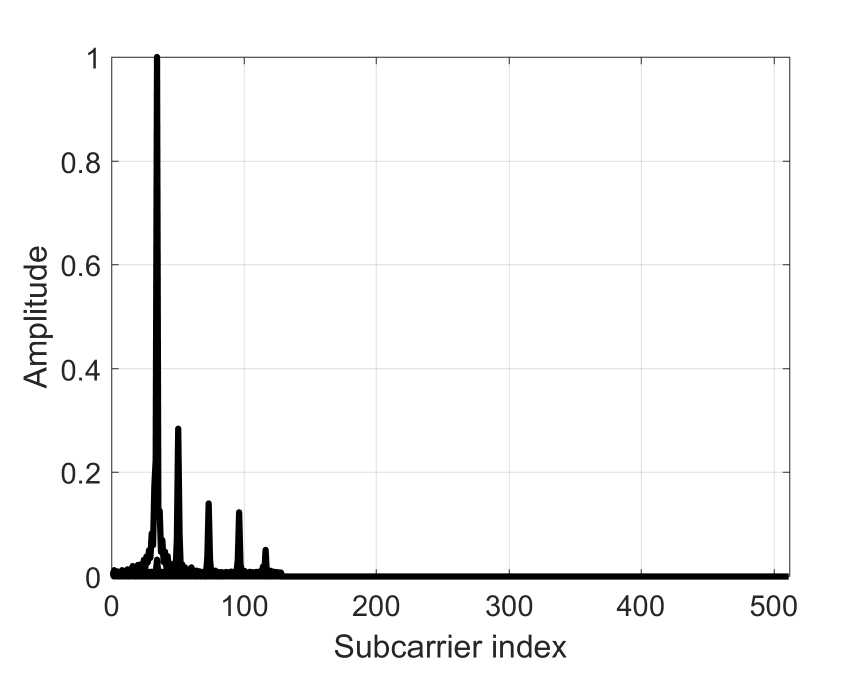}
\caption{Periodogram's normalized peaks for 5 targets, given~\gls{SNR} = 0 dB, $N$ = 512, and $N_{cp}$ = 128.}
\label{peaks_diagram}
\end{figure}
\section{Sensing Performance Enhancement}
\subsection{Channel Estimation Using \gls{DFT}-\gls{CE}}
As presented previously, since targets' velocities and ranges can be extracted from the estimated channel information, the overall detection precision depends on the quality of $\boldsymbol{H}$ predicted using (\ref{est_matrix}). The intuition is that an accurate estimation of $\boldsymbol{H}$ would output a precise parameter estimate. The approach that we propose below relies on the channel estimates using \gls{DFT}-\gls{CE}. \\
\indent It is worth mentioning that the~\gls{CP} is longer than the maximum channel delay. As such, each path delay of the multipath channel is lower than the time required to transmit the CP. More explicitly, in order to estimate ranges and velocities, instead of using all the channel impulse for the whole OFDM frame transmission time, we only use a short part of the channel impulse equivalent to the transmission time $T_{CP}$. At extremely low \gls{SNR}, the noise level is higher than many targets' peaks. Thus, the periodogram can pick up wrong peaks as targets with a high probability.
  \begin{figure*}[t!]
  \begin{minipage}{1\textwidth}
        \begin{subfigure}[b]{0.45\linewidth}
            \centering
            \includegraphics[width=\linewidth]{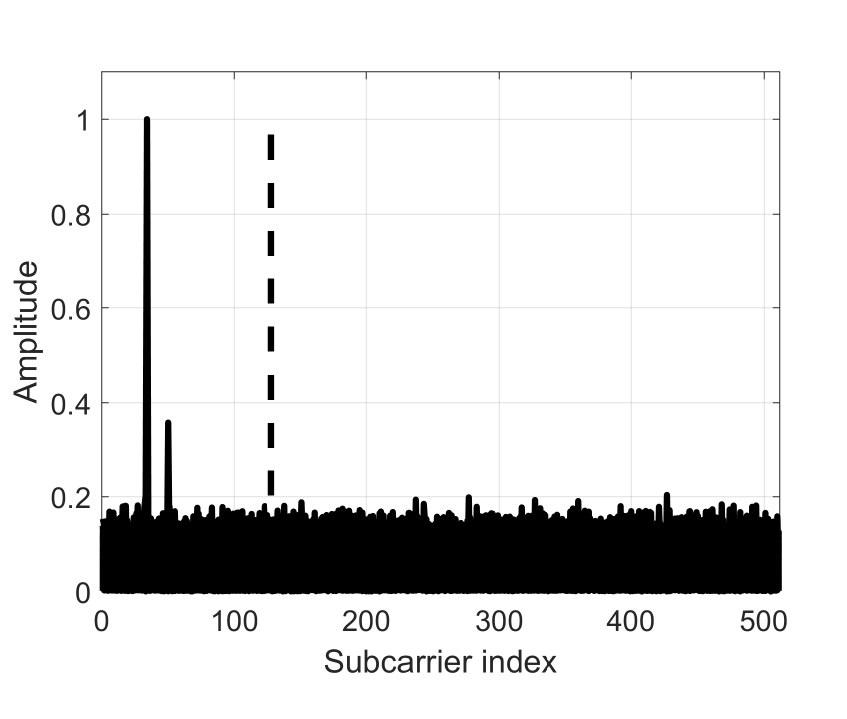}
            \caption{\gls{LS}-\gls{CE}.}
            \label{peak_level1}
        \end{subfigure}
        \hfill
        \begin{subfigure}[b]{0.45\textwidth}
            \centering
            \includegraphics[width=\linewidth]{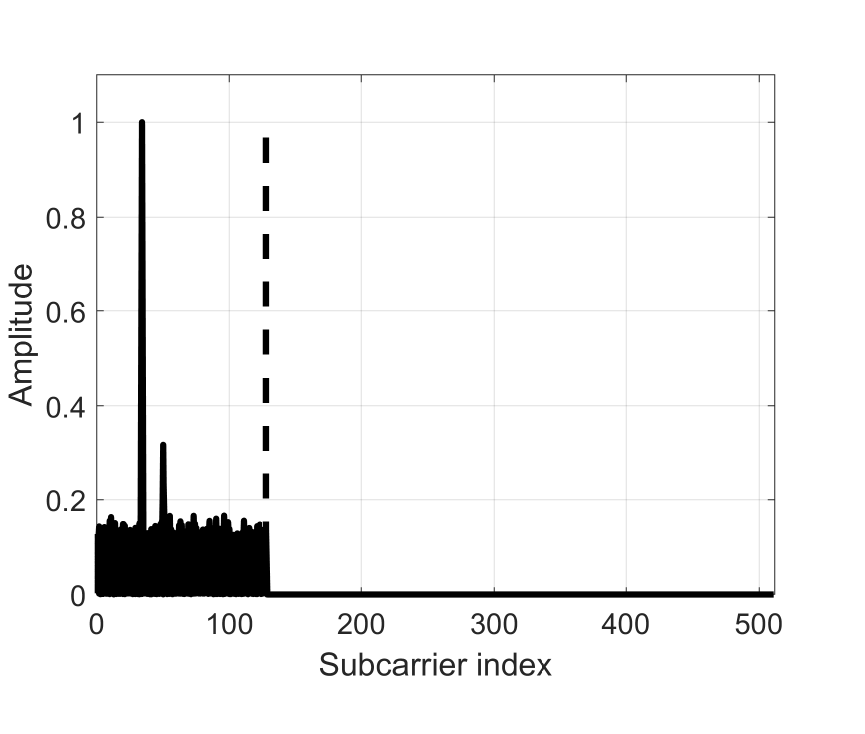}
            \caption{\gls{DFT}-\gls{CE}.}
            \label{peak_level2}
        \end{subfigure} 
        \caption{Normalized peak level as a function of the subcarrier index  for 512-\gls{OFDM} with $N_{cp}$ = 128 and~\gls{SNR} = -25dB.}
  \label{peak_level}
  \end{minipage}
  \end{figure*}
Using \mbox{\gls{DFT}-\gls{CE}} eliminates all the unuseful parts where targets are less likely to exist, which reduces the probability of the false detection caused by these peaks.\\
\indent The \gls{DFT}-\gls{CE} process can essentially be performed in 3 steps \mbox{\cite{5305570}:}
\begin{itemize}
    \item
    Transform the frequency-domain channel $\boldsymbol{H}$ to the time domain using \gls{IFFT}:
    \begin{equation}\label{IFFT}
    \begin{aligned}
        \hat{h}(n)=\frac{1}{K} \sum_{k=0}^{K-1} H(k) e^ {j 2 \pi \frac{n k}{K}}, ~  
      & 0 \leq n \leq K-1.
        \end{aligned}
    \end{equation}
    \item Restrain the effect of noise by keeping only the cyclic prefix equivalent part of the signal:
    \begin{equation}\label{Signal_rest}
        \hat{h}_r(n)=\left\{\begin{array}{cc}
        \hat{h}(n), & 0 \leq n \leq N_{cp}-1 \\
        0, & \text { otherwise }
        \end{array}\right.
    \end{equation}
    \item  Convert the channel estimate back to the frequency domain using~\gls{FFT}:
    
    \begin{equation}\label{FFT}
        \begin{aligned}
            \hat{H}^{\prime}(k)=& \frac{1}{K} \sum_{k=0}^{K-1} \hat{h}_r(n) e^{-j 2 \pi \frac{n k}{K}}, ~
            & 0 \leq n \leq K-1.
            \end{aligned}
    \end{equation}
\end{itemize}
The operations of converting the frequency-domain channel to the time domain, then the time-domain channel back to the frequency-domain counterpart are fast since the~\gls{IDFT} and~\gls{DFT} are implemented through~\gls{IFFT} and~\gls{FFT}, respectively. This improves the efficiency of channel estimation with low complexity.\\
\indent In order to obtain accurate channel state information, a better approach is to use \mbox{\gls{MMSE}-\gls{CE}}, which performs better than both~\gls{LS}-\gls{CE} and~\gls{DFT}-\gls{CE}. However,~\gls{MMSE}-\gls{CE} is limited by its high complexity and also by the fact that we need the real-time channel statistics (such as the co-variance matrix), which are hard to know in practice. Even though in~\cite{5305570} the authors proposed~\gls{MMSE} based \gls{DFT}-\gls{CE}, which is faster than the~\gls{MMSE}-\gls{CE}, its complexity remains very high and prohibitive for some applications when handling large matrices.\\
\indent Fig.~\ref{MSE_comparison} depicts the estimated \gls{MSE} of~\mbox{\gls{LS}-\gls{CE}},~\gls{LMMSE} based~\mbox{\gls{LS}-\gls{CE}},~\mbox{\gls{DFT}-\gls{CE}}, and~\gls{LMMSE} based~\mbox{\gls{DFT}-\gls{CE}}. As shown in the figure, it is clear that~\gls{LMMSE} based \mbox{\gls{DFT}-\gls{CE}} outperforms the others followed by \mbox{\gls{DFT}-\gls{CE}}. In contrast, \mbox{\gls{DFT}-\gls{CE}} provides a good trade-off between desirable estimation performance (compared to \gls{LS}-\gls{CE}) and low complexity (compared to~\gls{LMMSE} based~\gls{DFT}-\gls{CE}).\\
\indent Fig.~\ref{peaks_diagram} and Fig.~\ref{peak_level}  present the levels of peaks according to~\gls{OFDM} subcarriers index for a five-target case. When~\gls{SNR} is high, all the five targets are clearly detectable as can be seen in Fig. \ref{peaks_diagram}. However, knowing that targets can be located at subcarrier indexes $i$ such that $0 \le i \le N_{cp}-1$, when~\gls{SNR} becomes very low, for instance~\gls{SNR} = -25 dB as presented in Fig.~\ref{peak_level}, some or all targets peaks are confused with the noise. Subsequently, they can be incorrectly selected in subcarrier index $i$, such that $i \ge N_{cp}-1$, which increases the detection errors as shown in Fig. \ref{peak_level1}. By applying \gls{DFT}-\gls{CE}, however, targets can still be erroneously selected in the right subcarrier index range, i.e., $0 \le i \le N_{cp}-1$, hence reducing the estimation errors as depicted in Fig. \ref{peak_level2}.

\subsection{Zadoff-Chu Precoding}
\begin{figure}[t!] 
\centering
\includegraphics[width=3.5in]{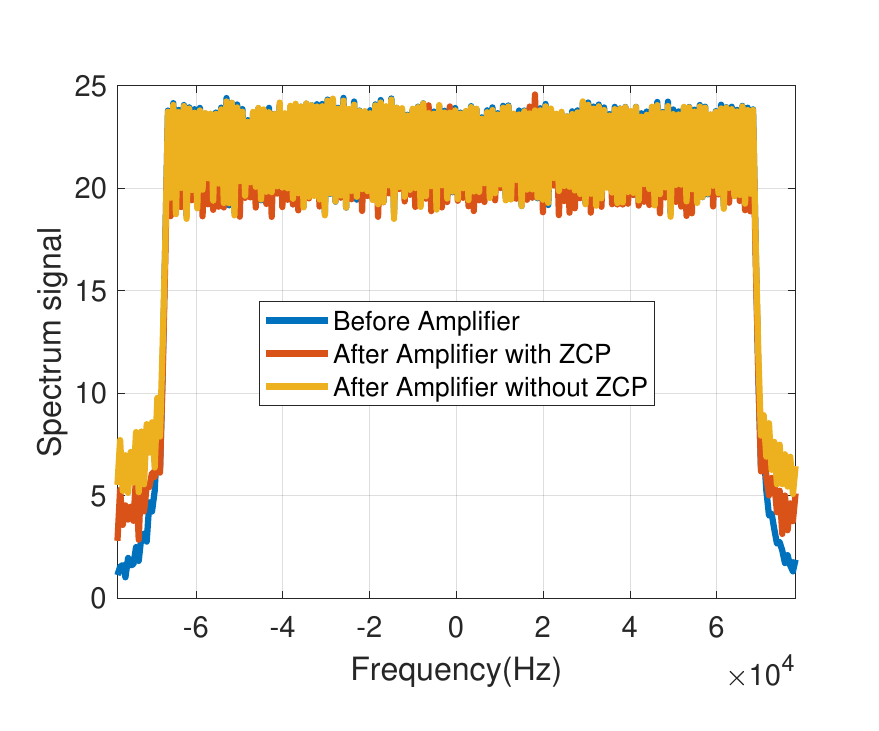}
\caption{Spectrum of a single \gls{QAM} symbol as a function of the frequency.}
\label{spectre}
\end{figure}

Zadoff-Chu sequences are known as a polyphase complex valued sequences with constant amplitude zero auto-correlation waveform (CAZAC). Because of their good  auto-correlation properties, these sequences have many applications especially in  3GPP Long Term Evolution (LTE) for synchronization of mobile phones with base stations.\\
A Zadoff-Chu sequence of length $L$ can be defined as: 
\begin{equation}
Z_{seq}(k')= \begin{cases}e^{\frac{j 2 \pi r'}{L}\left(\frac{k'^{2}}{2}+q' k'\right)}, & \text { when $L$ is even } \\ e^{\frac{j 2 \pi r'}{L}\left(\frac{k'(k'+1)}{2}+q' k'\right)}, & \text { when $L$ is odd }\end{cases}
,\end{equation}
where $k' = 0, 1,...,L-1$; $q' \in \mathbb{Z}$; $r'$ is an arbitrary integer relatively prime to $L$. From a Zadoff-Chu sequence of length $L$, we construct a square Zadoff-Chu matrix $\boldsymbol{Z_m}$ such that: 
\begin{equation}
    \begin{aligned}
     Z_{seq}(k')=Z_{seq}(iD+j)=Z_m(i,j),\\ 
     0\le i,j \le D-1 , D=\sqrt{L}.
     \end{aligned}
\end{equation}
At the transmitter, we precode the \gls{QAM} symbols by multiplying them with the matrix $\boldsymbol{Z_m}$. The precoded signal $\boldsymbol{A_p}$ is therefore $\boldsymbol{A_p}$=$\boldsymbol{Z_m}\boldsymbol{A}$.  After the \gls{IFFT}, the \gls{PAPR} of the signal is expressed as
\begin{equation}
PAPR=\frac{\max \left|x_n\right|^2}{E\left[\left|x_n\right|^2\right]}.
\end{equation}
At the receiver, the precoding can be discarded by multiplying the precoded received signal by $\boldsymbol{(Z_m})^{-1}$.
One of the major drawbacks of~\gls{OFDM} is its high~\gls{PAPR}. An~\gls{OFDM} signal with high~\gls{PAPR} is highly sensitive to nonlinear distortion caused by an~\gls{HPA}. This distortion increases the \gls{ACI} \cite{5616509,amhoud2017experimental}. Therefore~\gls{PAPR} reduction techniques may be employed to help cut it down. As demonstrated in~\cite{5616509}, \gls{ZCP} is used as a~\gls{PAPR} shrinking technique.\\
\indent In Fig.~\ref{spectre}, we compute the spectrum of one~\gls{QAM} symbol in the \mbox{128-\gls{OFDM}} signal with \mbox{$B$ = 20 MHz}, with an \gls{OFDM} signal going through an \gls{HPA} with an output:
\begin{equation}
 f_x=\frac{|x|}{\sqrt{ (1+(|x|/(\overline{x}\times 10^{q/10}))^2)}} ,
\end{equation}
 with state level $q$ = 2; $|x|$  and $\overline{x}$ are the modulus and the mean of $x$, respectively; \mbox{$\Delta f=\frac{B}{128}$}. Here, each data symbol is within $[ -\Delta f/2 , \Delta f/2]$.
 As a result, the ACI of each modulated symbol in the OFDM  based \gls{ZCP}, represented by the red curve is low compared to the \gls{ACI} of the OFDM without \gls{ZCP} illustrated by the yellow curve.  
In this work, we aim to analyze the impact of~\gls{ZCP} on the range and velocity estimations. By applying the previous~\gls{ZCP} on the~\gls{OFDM} signal, we end up with a signal with lower~\gls{PAPR}, lower \gls{ACI}, and improve the range and velocity estimation. 
\subsection{Complexity Reduction}
\begin{figure}[t!] 
\centering
\includegraphics[width=2.9in]{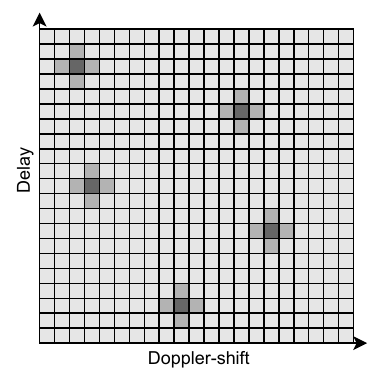}
\caption{2D sparse signal: The entire block is the signal frame, each slot represents a potential target. The dark gray  slots represent the effective targets.}
\label{sparse_signal}
\end{figure}

The~\gls{DFT} is one of the most important and widely used algorithms in computational tasks, such as in signal processing, communications, and audio/image/video compression. It is often implemented through~\gls{FFT} which computes the DFT of an $n$-dimensional signal with complexity $\mathcal{O}(n\log n)$.~\gls{FFT} does not make any assumption about the structure of the signal. However, in many applications, the signal is highly sparse in the frequency domain as depicted in Fig.~\ref{sparse_signal}. A signal $x$ is exactly \mbox{$K$}-sparse (or approximately \mbox{$K$}-sparse) in the frequency domain if its Fourier transform contains exactly $K$ non-zero values and the others are zeros. That is, the Fourier transform contains only $K$ dominant values and the others are close to zero. In radar applications, only a few targets are of interest, such as buildings, trucks, walls, and, thus, the final output is often sparse.\\
\indent Due to the importance of processing sparse data in many areas, several algorithms have been proposed to reduce not only the sampling complexity but also the computational complexity of \gls{FFT}. A detailed study of such algorithms can be found in \cite{4472244,8519339,inbooknearly,6736670}. Some of these algorithms can even reach a computational complexity of $\mathcal{O}(K \log K)$ for exactly \mbox{$K$-sparse} signals. The basic principle of all \gls{SFT} algorithms is to reduce the number of involved signal samples and therefore the computational complexity. Thus, the significant frequencies in the signal are first localized and then estimated, either iteratively or simultaneously.
However, most of these algorithms are only proposed for one-dimensional signals, and the extension of these algorithms to multi-dimensional signals is not straightforward. When dealing with multi-dimensional signals, it is required to create a large one-dimensional vector containing the entire signal, then after the processing step, to return to the initial dimensions to determine the index of dominant frequencies. To alleviate many of the drawbacks brought by the aforementioned techniques, in~\cite{8519339}, Wang et al. proposed \mbox{\gls{FPS}-\gls{SFT}}, which is a multi-dimensional and iterative \gls{SFT} algorithm. It processes a multi-dimensional sparse signal with low complexity and small samples under both noise-free and noisy conditions. Consequently, since the signals interested in this paper are highly sparse (\mbox{$K\ll N\times M$}), in this subsection, we take a critical look at \gls{FPS}-\gls{SFT} algorithm  in terms of its complexity and accuracy. The details of the algorithm can be found in~\cite{8519339}.
Since \gls{FPS}-\gls{SFT} is iterative and the frequencies recovered in a given iteration are passed to the next iteration, in low \gls{SNR} region, a frequency recovery error caused in a given iteration is carried through the next ones.
Let us assume that the \gls{FPS}-\gls{SFT} algorithm executes $I$ iterations and $Q$ is the least common multiple of the signal dimensions $N$ and $M$. For a general case of 2D signals, each iteration uses $3Q$ samples, since it is required that 3 $Q$-length slices are extracted to decode the two frequency components of a 2D signal in the frequency domain. Thus,  the  sampling  complexity  of  \gls{FPS}-\gls{SFT}  is $\mathcal{O}(3IQ) = \mathcal{O}(IQ)$. The core processing of \gls{FPS}-\gls{SFT} is the $Q$-point single-dimensional \gls{DFT}, which can be implemented by the \gls{FFT} with the computational  complexity  of $\mathcal{O}(Q\log Q)$. In addition to \gls{FFT}, each iteration needs to  evaluate  up  to $Q$ samples corresponding to different frequencies.  Hence,  the  computational complexity of~\gls{FPS}-\gls{SFT} is \mbox{$\mathcal{O}(I(Q\log Q+Q))=\mathcal{O}(IQ\log Q)$}. If we let the iteration size $I$ equal $I_{max}$ which is  sufficiently large so that \gls{FPS}-\gls{SFT} converges for a  given  $K$-sparse signal, the  sample and the computational complexity of~\gls{FPS}-\gls{SFT} become $\mathcal{O}(Q)$ and $\mathcal{O}(Q\log{}Q)$, respectively. For $K=\mathcal{O}(Q)$, \gls{FPS}-\gls{SFT}  achieves the lowest  sample and computational  complexity, i.e., $\mathcal{O}(K)$ and $\mathcal{O}(K\log K)$,  respectively, among all considered~\gls{SFT} algorithms~\cite{8519339}.
\vspace{0 cm}
 \begin{figure}[t!] 
\centering
\includegraphics[width=3.5in]{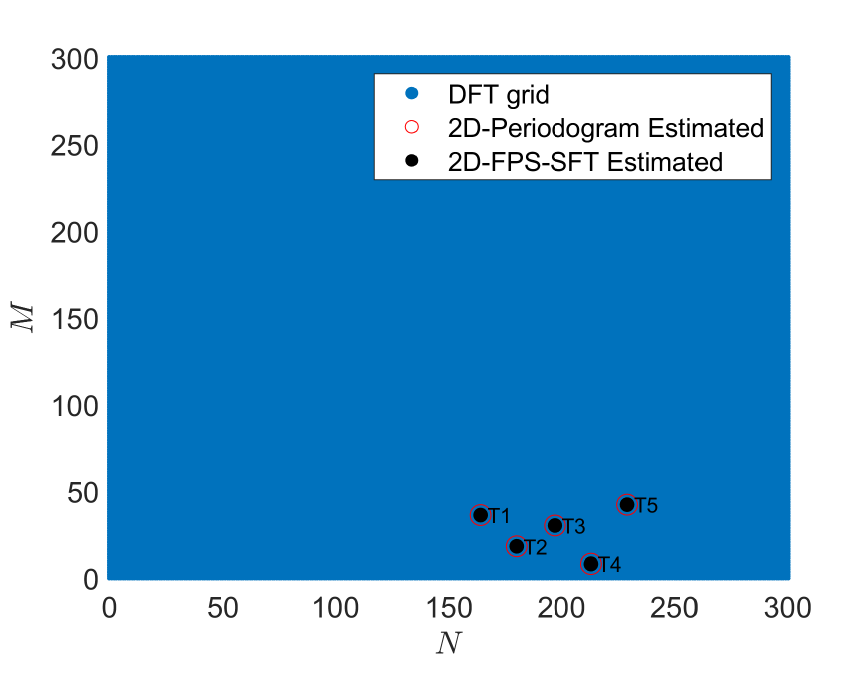}
\caption{2D periodogram and 2D-\gls{FPS}-\gls{SFT} estimate comparison for 5 targets with a 2048-\gls{OFDM}, $M=560$. Each target identified by its coordinates $(n,m)$.} 
\label{2D-FFT_vs_2D-SFT_estimated}
\end{figure}

 \begin{figure}[t!] 
\centering
\includegraphics[width=3.5in]{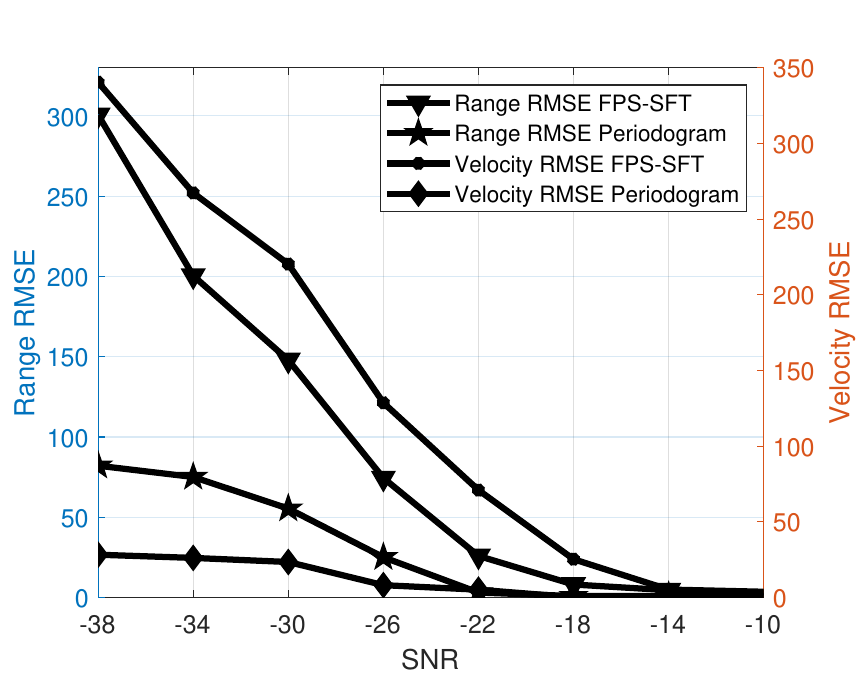}
\caption{2D-Periodogram and 2D-\gls{FPS}-\gls{SFT} estimate \gls{RMSE} as function of the SNR.}
\label{RMSE_PER_SFT}
\end{figure} 

\section{Simulation Results and Discussion}
\begin{figure}[t!] 
\centering
\includegraphics[width=3.5in]{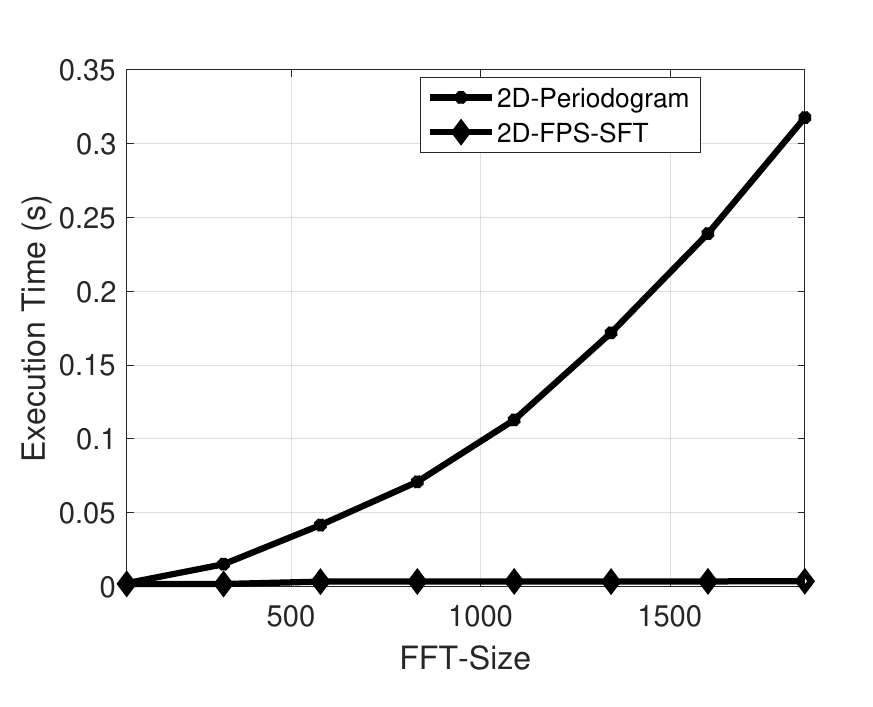}
\caption{2D periodogram and 2D-\gls{FPS}-\gls{SFT} execution time comparison with 5 targets versus the \gls{OFDM} subcarrier number size, given \mbox{$B$ = 60 MHz} and \mbox{$M$ = 200}.}
\label{2D_FFT_vs_2D_SFT_Execution_time}
\end{figure}
\begin{figure}[ht] 
\centering
\includegraphics[width=3.5in]{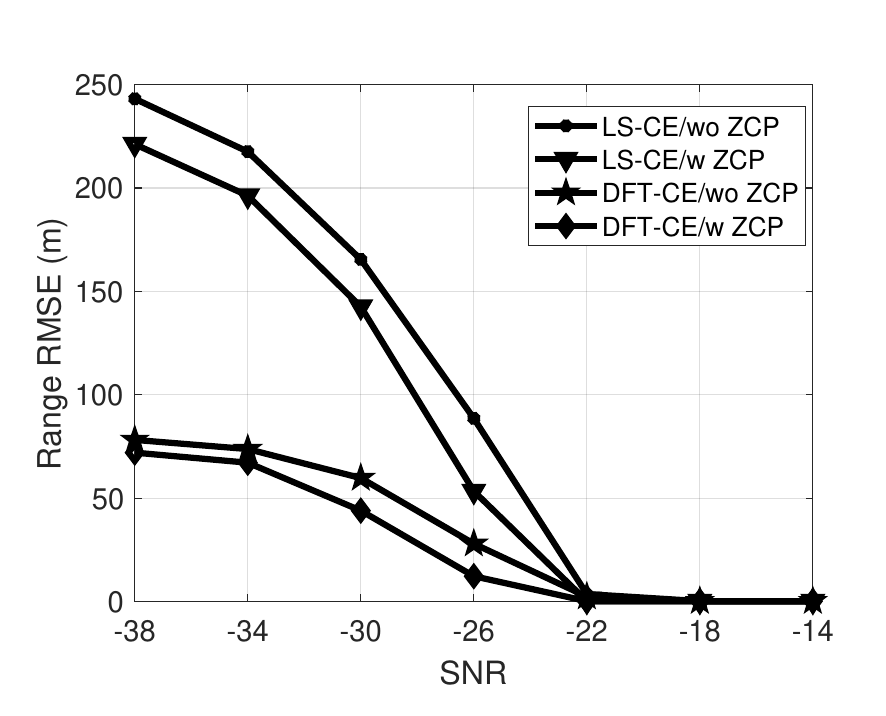}
\caption{Range \gls{RMSE} as a function of~\gls{SNR} for ZCP based LS-CE, LS-CE and DFT-CE.\\}
\label{RMSEd}
\end{figure}
\begin{table}[ht]
\caption{Simulation parameters}
\centering
\begin{tabular}{c c c }
\hline\hline
Parameter & Symbol & Value \\ [0.5ex] 
\hline
Number of OFDM symbols&$M$&560\\
Number of subcarriers&$N$&2,048\\
False alarm probability &$P_{fa}$&$10^{-2}$\\
Carrier frequency &$f_c$&77 GHz \\
Light speed&$c$&$3\times 10^8$ m/s \\
QAM-Size&$-$&16\\
Subcarrier space&$\Delta_f$&240 KHz \\
Bandwidth&$B$&491.52 MHz\\
Slots per subframe&$-$&4\\
Maximum range&$d_{max}$&156.25 m\\
Symbols per slot&$-$&14\\
Range resolution &$\Delta d$&30.52 cm\\
Velocity resolution &$\Delta v$&0.67 m/s \\
Cyclic prefix & $N_{cp}$ & 512  \\[1ex]
\hline
\end{tabular}
\label{simulation_parameters}
\end{table}
We evaluate the performance of the radar estimate in the \gls{ISM} band of 77~GHz for a total bandwidth of $B$ = 491,52~MHz using the proposed algorithms. The simulation parameters are inspired from the 5G New Radio (NR) specifications~\cite{5GNRUE}, and are summarized in  \mbox{Table \ref{simulation_parameters}}. In 5G NR, each frame duration is 10~ms, corresponding to 10 subframes. The 240~KHz subcarrier spacing configuration contains 16 slots per subframe and 14~\gls{OFDM} symbols by slot. Consequenty, this gives a total of 2,240 OFDM symbols per frame with \mbox{$T_{cp} = 7\% \times T$} (normal \gls{CP}). This \gls{CP} length is very small and can allow the radar to achieve only up to the coverage of \mbox{$d_{max}$ = 43.73~m}. For the simplicity of our simulations, we stipulate 4 slots per subframe along with a \gls{CP} such that \mbox{$T_{cp} = 25\% \times T$}, which results in a total of 560 OFDM symbols per frame and \mbox{$d_{max}$ = 156.25~m}. In our simulations, we are assuming that all $N$ \gls{OFDM} subcarriers carry useful information except the first \gls{OFDM} symbol of each frame, which contains block-type pilots for channel estimation at the communication receiver. We also assume that the multipath channel has sufficiently long coherence time, which makes it unchanged during the transmission of an \gls{OFDM} frame. The radar cross section is assumed to be unity. For the sake of simplicity, we further set $N'$ and $M'$ to be $N$ and $M$, respectively. With \mbox{$N$ = 2,048} and $M$ = 560. A Hamming window is used as the window function in (\ref{periodogram}). This configuration allows us to achieve a range resolution of 30.52 cm and a velocity resolution of 0.67 m/s. Such a radar can distinguish very close targets in terms of ranges  and velocities. Thus, a good parameterization of the radar and wave characteristics is crucial to achieve good target detection performance.

\begin{figure}[ht] 
\centering
\includegraphics[width=3.5in]{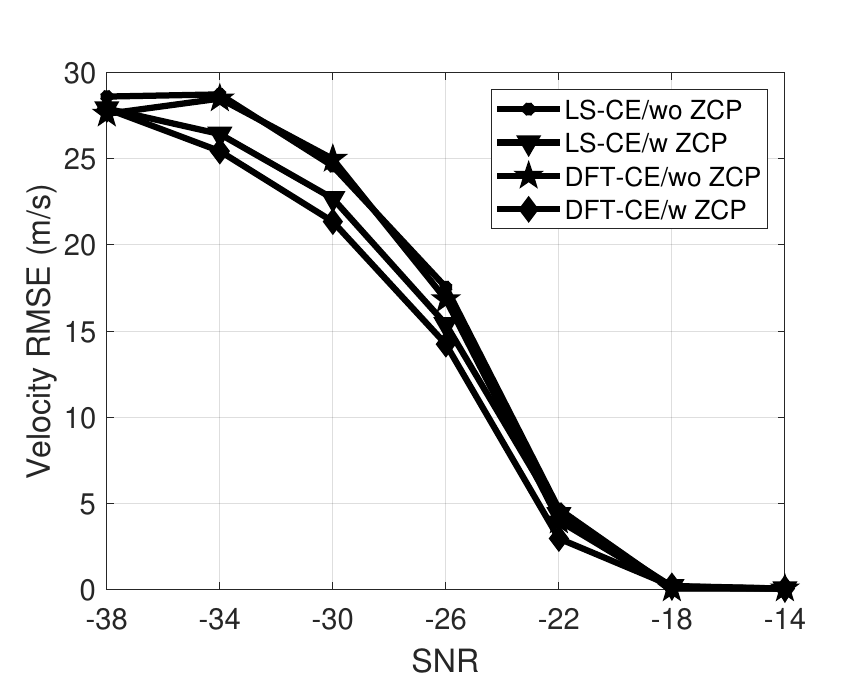}
\caption{Velocity \gls{RMSE} as a function of~\gls{SNR} for ZCP based LS-CE, LS-CE and DFT-CE.}
\label{RMSEv}
\end{figure}

\indent Fig. \ref{2D-FFT_vs_2D-SFT_estimated} shows the ordinary periodogram and the \gls{FPS}-\gls{SFT} estimates of a five-sparse-signal case at \mbox{\gls{SNR} = 20 dB}. We plot only the useful part of the signal where targets are located. From Fig. \ref{2D-FFT_vs_2D-SFT_estimated}, we can clearly observe that \gls{FPS}-\gls{SFT} and periodogram estimates perfectly match each other, which means that \gls{FPS}-\gls{SFT} can correctly estimate the dominant frequencies.\\
\indent Fig. \ref{RMSE_PER_SFT} depicts the \gls{RMSE} of the targets' range and velocity estimation using \gls{FPS}-\gls{SFT} and \gls{2D} periodogram. From the figure, we observe that 2D periodogram is more accurate than \gls{FPS}-\gls{SFT} in low SNR region. However, both methods converge in the high SNR region.\\
\indent Fig. \ref{2D_FFT_vs_2D_SFT_Execution_time} shows the execution time comparison of \mbox{\gls{FPS}-\gls{SFT}} and the periodogram. From the figure, we can notice that the periodogram execution time depends on the \gls{OFDM} size, whereas \mbox{\gls{FPS}-\gls{SFT}} is almost constant only depending on the sparsity order $K$ of the signal while drastically reducing the execution time.\\
\indent From Fig. \ref{RMSE_PER_SFT} and Fig. \ref{2D_FFT_vs_2D_SFT_Execution_time}, we conclude that a compromise is to be taken between the accuracy and the complexity of both 2D periodogram and  \gls{FPS}-\gls{SFT} depending on the type of application. \\
 \indent Fig. \ref{RMSEd} and Fig. \ref{RMSEv} show the \gls{RMSE} of the range and velocity estimation for the standard periodogram using \gls{LS}-\gls{CE} and \gls{DFT}-\gls{CE} with and without \gls{ZCP}. From these figures, we can observe that \gls{LS}-\gls{CE}, and \gls{DFT}-\gls{CE} using \gls{ZCP} outperforms the case without \gls{ZCP}. Moreover, applying \gls{DFT}-\gls{CE} based \gls{ZCP} yields better estimates than \gls{LS}-\gls{CE} based \gls{ZCP}. However, all the algorithms converge in higher~\gls{SNR} regions.
\section{Conclusion}
In this work, we proposed an improved target detection method using several emergent techniques for the periodogram estimation proceeding. The periodogram algorithm outputs estimation in the limit of the radar resolution which depends on the signal parameterization. Through our simulations, we established that when dealing with low~\gls{SNR}s, the estimation using \gls{LS}-\gls{CE} performs poorly since the targets' peaks are confused with noise. By using \gls{DFT}-\gls{CE}, we improved the channel estimation and reduced the target estimation error by filtering some false positive targets. Moreover, we concluded that \gls{ZCP}, in addition to its important applications in \gls{OFDM} transmission for \gls{PAPR} reduction and \gls{ICI} alleviation,  can also improve  target estimation performance in low~\gls{SNR} region. Considering that received signals are often sparse, \mbox{\gls{FPS}-\gls{SFT}} effectively reduces the estimation complexity but has poor accuracy in low \gls{SNR} region. In our future works, we aim to apply a noise reduction method to Zadoff-Chu precoded signals in order to provide more accurate channel estimates at low~\gls{SNR}. An extended version of those algorithms could be applied to networks of unmanned aerial vehicles where targets are moving in three dimensions. Furthermore, the overall estimation performance can also be improved using machine learning, which could be a worthwhile research direction in the future.
\balance
\bibliographystyle{IEEEtran}
\bibliography{biblio_traps_dynamics}

\begin{IEEEbiography}[{\includegraphics[width=1in,height=1.25in,clip,keepaspectratio]{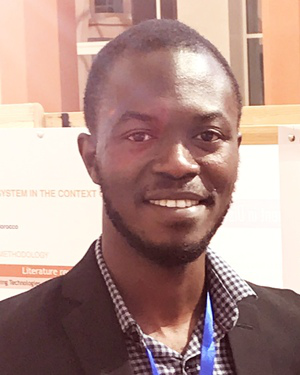}}]{MAMADY DELAMOU} received the M.S. degree
in telecommunication and networking from the National Institute of Posts and Telecommunications, Rabat, in 2019. After his Pre-doc, he is currently conducting his PhD with the School of Computer Science,  Mohammed VI Polytechnic University (UM6P), Morocco. His research interests are wireless and wired communication, signal processing, radar detection, and machine learning.
\end{IEEEbiography}
\begin{IEEEbiography}[{\includegraphics[width=1in,height=1.25in,clip,keepaspectratio]{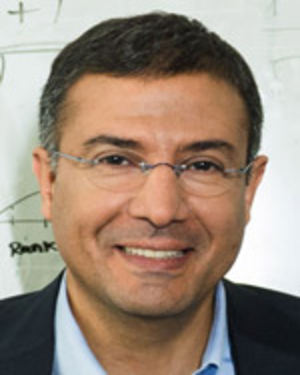}}]{Guevara Noubir} holds a PhD in Computer Science from the Swiss Federal Institute of Technology in Lausanne (EPFL) (1996). His research covers both theoretical and practical aspects of privacy, security, and robustness in networked systems. Prior to joining Northeastern University, he was a senior researcher at CSEM SA (1997-2000) where he led the design and development of the data protocol-stack of the third generation Universal Mobile Telecommunication System (UMTS) and its world-first 3G prototype. His research led to a wide range of mechanisms and algorithms for scalable, secure, private, and robust wireless and mobile communications. He led the winning team of the 2013 DARPA Spectrum Cooperative Challenge against 90 academic and industry teams. He is a recipient of the National Science Foundation CAREER Award (2005), the ACM Conference on Security and Privacy in Wireless and Mobile Networks (WiSec) best paper award in 2011 and runner-up best paper in 2013. His research was featured in the NSF CISE/CNS Highlights in 2009 and 2012. Professor Noubir has held visiting research positions at Eurecom, MIT, and UNL.
Professor Noubir has served as program co-chair of many conferences in his areas of expertise, including the ACM Conference on Security and Privacy in Wireless and Mobile Networks, IEEE Conference on Communications and Network Security, and IEEE WoWMoM. He also co-chaired two NSF workshops on bio-computation and communications. He serves on the editorial board of the IEEE Transaction on Mobile Computing, the Elsevier Journal on Computer Networks, and the ACM Transactions on Information and System Security.
\end{IEEEbiography}
\begin{IEEEbiography}[{\includegraphics[width=1in,height=1.25in,clip,keepaspectratio]{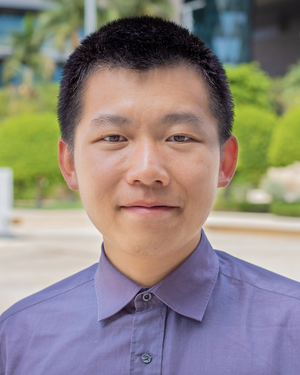}}]{SHUPING DANG} (S’13–M’18) received the B.Eng (Hons) in Electrical and Electronic Engineering from the University of Manchester (with first class honors) and B.Eng in Electrical Engineering and Automation from Beijing Jiaotong University in 2014 via a joint ‘2+2’ dual-degree program. He also received D.Phil in Engineering Science from University of Oxford in 2018. Dr. Dang joined in the R\&D Center, Huanan Communication Co., Ltd. after graduating from University of Oxford and then working as a Postdoctoral Fellow with the Computer, Electrical and Mathematical Science and Engineering Division, King Abdullah University of Science and Technology (KAUST). He is currently a Lecturer with Department of Electrical and Electronic Engineering, University of Bristol. The research interests of Dr. Dang include 6G communications, wireless communications, wireless security, and machine learning for communications.
\end{IEEEbiography}
\begin{IEEEbiography}[{\includegraphics[width=1in,height=1.25in,clip,keepaspectratio]{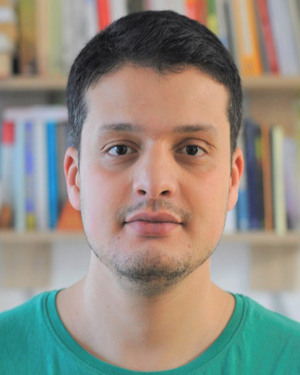}}]{EL MEHDI AMHOUD} (Member, IEEE) received
the Ph.D. degree in computer and communication
sciences from Télécom ParisTech, France. He is
currently an Assistant Professor with Mohammed
VI Polytechnic University, Morocco. Prior to his
current position, he was a Postdoctoral Research
Fellow with the King Abdullah University of Science and Technology, Saudi Arabia. He holds several U.S. patents. His research interests include modeling and analyzing the performance of new generations of communication networks and the Internet of Things. He received two awards of excellence for his outstanding Ph.D. thesis from the Mines-Télécom Institute and the Marie Skłodowska-Curie Research Grant from the European Commission.
\end{IEEEbiography}
\EOD
\end{document}